\pdfoutput=1
\documentclass[12pt,titlepage]{article}

\font\grande=cmr9.5 scaled \magstep4
\font\medio=cmr9.5 scaled \magstep2
\outer\def\beginsection#1\par{\medbreak\bigskip
      \message{#1}\leftline{\bf#1}\nobreak\medskip
\vskip-\parskip
      \noindent}
\usepackage{graphicx} 
\usepackage{mathrsfs}
\begin{document}

\bibliographystyle{unsrt}

\titlepage

\vspace{1cm}
\begin{center}
{\grande Relic gravitons from stiff curvature perturbations}\\ 
\vspace{1cm}
Massimo Giovannini \footnote{e-mail address: massimo.giovannini@cern.ch}\\
\vspace{1cm}
{{\sl Department of Physics, CERN, 1211 Geneva 23, Switzerland }}\\
\vspace{0.5cm}
{{\sl INFN, Section of Milan-Bicocca, 20126 Milan, Italy}}
\vspace*{1cm}
\end{center}
\vskip 0.3cm
\centerline{\medio  Abstract}
The tensor modes reentering the Hubble radius when the plasma is dominated by a stiff fluid lead to a spectral energy density whose blue slope depends on the total post-inflationary sound speed. This result gets however corrected  by a secondary (gauge-dependent) 
term coming from the curvature inhomogeneities that reenter all along the same stage of expansion. In comparison with the first-order result, the secondary contribution is shown to be always suppressed inside the sound horizon  and its effect  on the total spectral energy density of the relic gravitons is therefore negligible for all phenomenological purposes. It is also suggested that the effective anisotropic stress of the curvature inhomogeneities can be obtained from the functional derivative of the second-order action of curvature inhomogeneities with respect to the background metric. 
\vskip 0.1cm
 \noindent
\vspace{5mm}
\vfill
\newpage

A primeval stiff and irrotational fluid has been originally proposed, with different motivations, by Zeldovich \cite{ONE}, Sakharov \cite{TWO} and Grishchuk \cite{THREE}. After the formulation of the conventional inflationary paradigm Ford \cite{FOUR} observed that gravitational particle production at the end of inflation could account for the entropy of the present universe and suggested that the backreaction effects of the created quanta constrain the length of a stiff post-inflationary phase by making the expansion dominated by radiation. It has been later argued by  Spokoiny \cite{FIVE} that various classes of scalar field potentials exhibit a transition from inflation to a stiff phase dominated by the kinetic energy of the inflaton.  A generic signature of a post-inflationary phase stiffer than radiation is the production of relic gravitons with increasing spectral energy density \cite{SIX}. 
In quintessential inflationary models the inflaton and the quintessence field are identified in a single scalar degree of freedom \cite{SEVEN,EIGHT} and various concrete forms of the inflaton-quintessence potential have been scrutinized through the years. The transition between an inflationary phase and a kinetic phase can be realized both with power-law potentials and with exponential potentials. Within the Palatini approach it has been recently suggested that  in the presence of a generalized gravitational action, the slow-roll parameters and the tensor-to-scalar ratio can be suppressed in comparison with the conventional situation \cite{EIGHTa,EIGHTb,EIGHTc,EIGHTd}. In this case the inflation-quintessence field can also have a power-law dependence during inflation \cite{EIGHTe} without conflicting with the standard bounds on the tensor-to-scalar ratio. In different frequency domains the spectra of the relic gravitons can be used not only to infer the early evolution of the space-time curvature, but also to test the evolution of the plasma for temperatures in the MeV range, i.e. prior to the formation of light nuclei \cite{EIGHTf}.  Different classes of models have been proposed and are overall consistent with the observational constraints \cite{EIGHTg,EIGHTh}. The occurrence of a stiff post-inflationary phase may even increase the maximal number of $e$-folds that are today accessible by large-scale observations \cite{SIX,EIGHTi}.

The power spectra of the tensor modes exiting the Hubble radius during the inflationary epoch and reentering in a generic stiff phase are less suppressed than in the case of a radiation-dominated evolution. More specifically, if the stiff scale factor evolves as $a(\tau) \simeq \tau^{\alpha}$  in the conformal time parametrization the corresponding tensor power spectra are:
\begin{equation}
P^{(st)}_{T}(q,\tau) = \overline{P}_{T}(q) \, \frac{\sin^2{q\tau}}{|q \,\tau|^{2 \alpha}}, \qquad \overline{P}_{T}(q) = {\mathcal A}_{T} \biggl(\frac{q}{q_{p}}\biggr)^{n_{T}},\qquad \alpha = \frac{2}{3 w+1 },
\label{EQ1}
\end{equation}
where the superscript reminds that the corresponding modes exited during inflation but reentered in the stiff epoch; following the standard notations $w$ is the barotropic index; ${\mathcal A}_{T}$ denotes the amplitude of the tensor power spectrum at the pivot scale $q_{p}$ and $n_{T}$ is the tensor spectral index. If $1/3 < w \leq 1$ the slope appearing in Eq. (\ref{EQ1}) is such that $1/2 \leq \alpha < 1$. 
The stiff power spectrum inside the Hubble radius (i.e. $q > a\, H$) is therefore enhanced in comparison with the analog result valid  during a radiation-dominated stage of expansion; using standard notations the Hubble rate will be denoted by $H$ while ${\mathcal H} = a \, H = a^{\prime}/a$ where the prime denotes a derivation with respect to the conformal time coordinate $\tau$. The spectrum of Eq. (\ref{EQ1}) can be related to its radiation counterpart as: 
\begin{equation}
P^{(st)}_{T}(q,\tau) = \biggl|\frac{q}{a \, H} \biggr|^{2 ( 1 - \alpha)} \,P^{(rad)}_{T}(q,\tau), \qquad P^{(rad)}_{T}(q,\tau) = 
\overline{P}_{T}(q) \, \frac{\sin^2{q \tau}}{| q\,\tau|^2},
\label{EQ2}
\end{equation}
where $P^{(rad)}_{T}(q,\tau)$ is the tensor power spectrum of the modes exiting during inflation and reentering 
when the background is dominated by radiation.  
All in all Eqs. (\ref{EQ1}) and (\ref{EQ2}) imply that the stiff power spectra
are comparatively less suppressed than their radiation analog  inside the Hubble radius, i.e. 
$P_{T}^{(st)}(q,\tau)  \gg P_{T}^{(rad)}(q,\tau)$ for $q \gg a\, H$.

In full analogy with the tensor modes, the stiff curvature inhomogeneities (denoted hereunder by ${\mathcal R}$) are also enhanced when the corresponding wavenumbers  are shorter than the sound horizon $r_{s}(\tau)$:
\begin{equation}
r_{s}(\tau) = \int_{0}^{\tau} c_{st}(\tau^{\prime}) \, d\tau^{\prime}, \qquad c^2_{st}(\tau) = \frac{p_{t}^{\prime}}{\rho_{t}^{\prime}},
\label{EQ3}
\end{equation}
where $c_{st}(\tau)$ will denote throughout the sound speed of the plasma. 
The stiff spectra of curvature perturbations exceed their radiation analog and follow a scaling law that is similar to Eq. (\ref{EQ2}) with the difference that the Hubble radius is now replaced by the sound horizon of Eq. (\ref{EQ3}): 
\begin{equation}
P^{(st)}_{{\mathcal R}}(q,\tau) = |q\, r_{s}(\tau)|^{2 ( 1 - \alpha)} P^{(rad)}_{{\mathcal R}}(q,\tau), \qquad |q \, r_{s}(\tau)| \gg 1.
\label{EQ4}
\end{equation}
The enhancement of the stiff spectra of curvature inhomogeneities 
naively implies a larger second-order contribution to the relic graviton background but this expectation will now be more carefully scrutinized. Suppose, in a preliminary perspective, that the inflationary phase is followed by a radiation-dominated stage of expansion; in this situation the spectral energy density consists  of a first-order contribution supplemented by a second-order correction \cite{NINE,TEN}:
\begin{equation}
h_{0}^2 \,\Omega_{gw}(q,\tau_{0}) = \frac{r_{T} {\mathcal A}_{\mathcal R}\,h_{0}^2\,\Omega_{R0}}{12} \biggl(\frac{q}{q_{p}}\biggr)^{n_{T}} \biggl[ 1 + 
\frac{{\mathcal A}_{{\mathcal R}}}{r_{T}} f(q, n_{s}) \biggl(\frac{q}{q_{p}}\biggr)^{\gamma}\biggr],\qquad 
\gamma = 2 (n_{s} -1) - n_{T},
\label{EQ5}
\end{equation}
where $n_{s}$ is the scalar spectral index and $f(n_{s}, q)$  is a slowly varying function of $n_{s}$ and $q$; $\Omega_{R0}$ will denote throughout the critical fraction of the energy density stored in the relativistic species (in the concordance paradigm $h_{0}^2 \Omega_{R0}= 4.15 \times 10^{-5}$).  In Eq. (\ref{EQ5}) the amplitude of curvature perturbations has been related to the tensor amplitude of Eq. (\ref{EQ1}) via the standard tensor to scalar ratio $r_{T}$ (i.e. ${\mathcal A}_{T} = r_{T} {\mathcal A}_{{\mathcal R}}$). For  $0.9< n_{s}  < 1$ (and in the whole range of comoving frequencies between the aHz and the MHz) $f(n_{s}, q)$ varies between $10^{-2}$ and $10^{-3}$. Equation (\ref{EQ5})  therefore suggests that the second-order contribution to the tensor background could be, in principle, rather large provided the scalar spectral index is blue as it happens, incidentally, in the case of waterfall transitions \cite{TEN,TENa}. Examples of equally large contributions have been proposed in the context of the reheating dynamics \cite{THIRTEENa,FOURTEEN}. 

While Eq. (\ref{EQ5}) holds in the case of a post-inflationary phase dominated by radiation, 
the first-order tensor mode functions reentering during the stiff epoch (i.e. $q\, \geq a \, H$) have a different form that can be computed within the Wentzel-Kramers-Brioullin (WKB) approximation:
\begin{equation}
h^{(1)}_{i\,j}(\vec{q},\tau) = \overline{h}_{i\,j}(\vec{q}\,) \,\frac{\sin{q\,\tau}}{|q\,\tau|^{\alpha}}, \qquad \partial_{\tau} h^{(1)}_{i\,j}(\vec{q},\tau) = q \,\, \overline{h}_{i\,j}(\vec{q}\,) \, \biggl[\frac{\cos{q \tau}}{|q \, \tau|^{\alpha}} - \alpha \frac{\sin{q\tau}}{|q\,\tau|^{\alpha+1}}\biggr],
\label{EQ6}
\end{equation}
where $\alpha$ has been already introduced in Eq. (\ref{EQ1}) and the superscript refers to the first-order contribution. In Eq. (\ref{EQ6})  $\overline{h}_{i\,j}(\vec{q}\,)$ is just a solenoidal, traceless and 
stationary random field whose correlation function obeys\footnote{For a gravitational wave propagating in the $\hat{q}$ direction the two orthogonal polarizations are  $e_{i\,j}^{(\oplus)}(\hat{q}) = [\hat{m}_{i} \,\hat{m}_{j} - \hat{n}_{i} \,\hat{n}_{j}]$ and $e_{i\,j}^{(\otimes)}(\hat{q}) = [\hat{m}_{i} \,\hat{n}_{j} + \hat{n}_{i} \,\hat{m}_{j}]$ 
where $\hat{m}$, $\hat{n}$ and $\hat{q}$ are three mutually orthogonal unit vectors; the sum over the two 
tensor polarizations is given, as usual, by $\sum_{\lambda} e^{(\lambda)}_{i\,j}(\hat{q}) e^{(\lambda)}_{m\,n}(\hat{q}) = 4 \, {\mathcal S}_{i\,j\,m\,n}(\hat{q})$ (see, for instance, \cite{EIGHTf}).}:
\begin{eqnarray}
\langle \overline{h}_{i\,j}(\vec{q}\,) \overline{h}_{m\, n} (\vec{q}^{\,\,\prime}) \rangle &=& \frac{2 \pi^2}{q^3} \, {\mathcal S}_{i\,j\,m\,n}(\hat{q}) \,\overline{P}_{T}(q) \, \delta^{(3)}(\vec{q} + \vec{q}^{\,\,\prime}),
\label{EQ7}\\
{\mathcal S}_{i\, j\, m\, n}(\hat{q}) &=&  \frac{1}{4} \biggl[ p_{im}(\hat{q}) p_{jn}(\hat{q}) + p_{in}(\hat{q}) p_{jm}(\hat{q}) - p_{ij}(\hat{q}) p_{mn}(\hat{q})\biggr].
\label{EQ8}
\end{eqnarray}
In Eq. (\ref{EQ8}) $p_{ij}(\hat{q}) =[ \delta_{ij} - \hat{q}_{i} \hat{q}_{j}]$ is the standard transverse projector and $\overline{P}_{T}(q)$ coincides with the expression of Eq. (\ref{EQ1}).  From Eq. (\ref{EQ6}) it is immediate to show, using Eq. (\ref{EQ8}) that 
\begin{equation}
\langle h^{(1)}_{i\,j}(\vec{q},\tau) \, h^{(1)}_{m\,n}(\vec{q}^{\,\,\prime},\tau) \rangle =  \frac{2 \pi^2}{q^3} \, {\mathcal S}_{i\,j\,m\,n}(\hat{q}) \,P_{T}(q,\tau) \, \delta^{(3)}(\vec{q} + \vec{q}^{\,\,\prime}).
\label{EQ8a}
\end{equation}
If the tensor modes reenter during a stiff stage of expansion the $P_{T}(q,\tau)$ appearing in Eq. (\ref{EQ8a}) coincides with $P^{(st)}_{T}(q,\tau)$ of Eq. (\ref{EQ1}).
The variation of the tensor action with respect to the background metric leads to the energy-momentum 
tensor of the relic gravitons whose corresponding energy density is therefore given by \cite{EIGHTf,FIFTEEN}:
\begin{equation}
\rho_{gw} = \frac{1}{8 \ell_{\mathrm{P}}^2 a^2} \biggl[ \partial_{\tau} h_{k \ell}\, \partial_{\tau}h^{k \ell} + \partial_{m} h_{k\ell} \partial^{m} h^{k\ell}\biggr],
\label{EQ9}
\end{equation}
where $h_{k\ell}$ is  given by the first-order contribution supplemented by the second-order terms that will be specified hereunder.  The random averages of Eqs. (\ref{EQ7}), (\ref{EQ8}) and (\ref{EQ8a})
can be used into Eq. (\ref{EQ9}) to obtain $\overline{\rho}_{gw}^{(1)} = \langle \, \rho_{gw} \, \rangle$;
consequently the first-order contribution to the spectral energy density becomes:
\begin{equation}
h_{0}^2 \,\Omega_{gw}^{(1)}(q,\tau_{0}) = \frac{h_{0}^2 }{\rho_{crit}} \frac{d \overline{\rho}^{(1)}_{gw}}{d \ln{q}} =
 \frac{h_{0}^2 \Omega_{R0}}{12} \, \overline{P}_{T}(q)  \biggl(\frac{q}{q_{r}}\biggr)^{2 ( 1 - \alpha)}, \qquad \rho_{crit} = \frac{ 3\,H^2 }{\ell_{P}^2},
 \label{EQ10}
 \end{equation}
 where $\ell_{P} = \sqrt{8 \pi G}$.
Equation  (\ref{EQ10}) encompasses all the modes that are inside the Hubble radius at the present time and for 
frequencies $q\,> \,q_{r} = a_{r}\, H_{r}$. The transition to radiation (occurring at the typical scale $H_{r}$) can be triggered by different kinds of reheating mechanisms \cite{FOUR,EIGHTg}; note that in  the case $\alpha = 1/2$ Eq. (\ref{EQ10}) inherits a phenomenologically relevant (logarithmic) contribution \cite{SEVEN,EIGHT} which will be neglected since it is not essential for the present ends.

The first-order spectral energy density of Eq. (\ref{EQ10}) is corrected by the effect of the curvature inhomogeneities reentering all along the stiff phase: this addition ultimately modifies both Eqs. (\ref{EQ2}) and (\ref{EQ10}).  To estimate this contribution in general terms the energy-momentum tensor of the stiff and irrotational fluid driving the post-inflationary evolution will be parametrized as $T_{\mu}^{\,\,\nu} = (\rho_{t} + p_{t}) u_{\mu} \, u^{\nu} - p_{t} \delta_{\mu}^{\nu}$ where $\rho_{t}$, $p_{t}$ and $u_{\mu}$ are the total energy density, the pressure and the four-velocity. 
During the post-inflationary phase the gauge-invariant curvature inhomogeneities corresponding to the normal modes of the irrotational fluid in a conformally flat background geometry evolve, in Fourier space, as \cite{NINEa}:
\begin{equation}
{\mathcal R}_{\vec{k}}^{\prime\prime} + 2 \frac{z_{t}^{\prime}}{z_{t}} {\mathcal R}_{\vec{k}}^{\prime} + k^2 \, c_{st}^2 {\mathcal R}_{\vec{k}} =0, \qquad z_{t} = \frac{a^2 \sqrt{ p_{t} + \rho_{t}}}{{\mathcal H} c_{st}}.
\label{EQ11}
\end{equation}
In the short-wavelength limit (i.e. $k^2 c_{s}^2 \gg |z_{t}^{\prime\prime}/z_{t}|$) the solution of Eq. (\ref{EQ11}) follows from the  WKB approximation
\begin{equation}
{\mathcal R}_{\vec{q}}(\tau) = \overline{{\mathcal R}}(\vec{q}\,) \frac{\sin{q\,r_{s}(\tau)}}{|q\,r_{s}(\tau)|^{\alpha}}, \qquad {\mathcal R}^{\prime}_{\vec{q}}(\tau) = q\, c_{st} \,\overline{{\mathcal R}}(\vec{q}\,)\biggl[\frac{\cos{q\,r_{s}(\tau)}}{|q \,r_{s}(\tau)|^{\alpha}} - \alpha \,\frac{\sin{q \,r_{s}(\tau)}}{|q \,r_{s}(\tau)|^{\alpha+1}}\biggr],
\label{EQ12}
\end{equation}
where  $\overline{{\mathcal R}}(\vec{q}\,)$ is a scalar (static) random field whose correlation function is similar to the one of Eq. (\ref{EQ7}) :
\begin{equation}
\langle \overline{{\mathcal R}}(\vec{q}\,) \, \overline{{\mathcal R}}(\vec{q}^{\,\prime}\,) \rangle = \frac{2 \pi^2}{q^3} \overline{P}_{{\mathcal R}}(q) \, \delta^{(3)}(\vec{q} + 
\vec{q}^{\,\prime}\,), \qquad \overline{P}_{{\mathcal R}}(q) = {\mathcal A}_{{\mathcal R}} \biggl(\frac{q}{q_{p}}\biggr)^{n_{s} -1},
\label{EQ13}
\end{equation}
where ${\mathcal A}_{{\mathcal R}}$ is the scalar amplitude at the pivot scale $q_{p}$ 
and  $n_{s}$ is the scalar spectral index [already introduced in Eq. (\ref{EQ5})].

If  ${\mathcal G}_{\mu}^{\,\,\,\,\nu}$ is the Einstein tensor and $T_{\mu}^{\,\,\,\,\nu}$ denotes a generic energy-momentum tensor of the matter sources,  the effective anisotropic stress 
determined by the curvature inhomogeneities can be formally expressed as:
 \begin{equation}
  \Pi_{i}^{\,\,\,j}(\vec{x}, \tau) =  \delta^{(2)}_{s} T_{i}^{\,\,\,j} - \frac{1}{\ell_{P}^2} \delta_{s}^{(2)} {\mathcal G}_{i}^{\,\,\,j},
 \label{EQ13a}
 \end{equation} 
 where $\delta_{s}^{(2)}$ is the second-order scalar fluctuations of the corresponding tensor. In a nutshell
Eq. (\ref{EQ13a}) summarizes the standard Landau-Lifshitz prescription \cite{FOURTEENa} for the estimate 
of the energy-momentum pseudo-tensor. Moreover
the explicit form of Eq. (\ref{EQ13a}) is clearly gauge-dependent  (see e.g. \cite{NINE} and also \cite{ELEVEN,TWELVE,THIRTEEN}). It can be shown, on a general ground, that the effective anisotropic stresses derived in different gauges coincide inside the sound horizon but differ significantly outside of it \cite{THIRTEEN}.  For the present estimates it will be sufficient to consider the expressions inside the sound horizon in any gauge. For instance the
 first-order perturbed entries of the metric in the longitudinal gauge are given by $\delta_{s}^{(1)} \, g_{00}= 2 \,a^2 \, \phi$ and $\delta^{(1)} g_{ij} = 2 \, a^2\, \psi\, \delta_{ij}$; using Eq. (\ref{EQ13a}).
a straightforward calculation leads to the following form of the effective anisotropic stress:
\begin{eqnarray}
\Pi_{i}^{\,\,\,j}(\vec{x}, \tau) &=& \frac{1}{ \ell_{P}^2 \, a^2} \biggl[ \partial_{i} \psi \partial^{j} \psi - \partial_{i} \phi \partial^{j} \phi  - \partial_{i} \phi \partial^{j} \psi  
- \partial_{i} \psi \partial^{j} \phi 
\nonumber\\
&+& 2 \psi \partial_{i} \partial^{j} ( \phi - \psi) - 2\,\frac{\partial_{i}(\psi^{\prime } + {\mathcal H} \phi) \partial^{j} (\psi^{\prime } + {\mathcal H} \phi) }{({\mathcal H}^2 - {\mathcal H}^{\prime})} 
\biggr].
\label{EQ13b}
\end{eqnarray}
In Fourier space Eq. (\ref{EQ13b}) can be directly expressed in terms of   ${\mathcal R}_{\vec{k}}(\tau)$ and ${\mathcal R}_{\vec{k}}^{\prime}(\tau)$ obeying Eq. (\ref{EQ11}):
\begin{eqnarray}
\Pi_{i\, j}(\vec{q}, \tau) &=& - \frac{2 ({\mathcal H}^2 - {\mathcal H}^{\prime})}{(2\pi)^{3/2}\,\ell_{P}^2\, a^2(\tau)\,{\mathcal H}^2 } \int\, d^{3}k\, \, k_{i} \,\, k_{j} \biggl\{ {\mathcal R}_{\vec{k}} \, {\mathcal R}_{\vec{q} - \vec{k}} + \frac{{\mathcal H}^2 - {\mathcal H}^{\prime}}{{\mathcal H}} \biggl[ \frac{{\mathcal R}_{\vec{k}} \, {\mathcal R}_{\vec{q} - \vec{k}}^{\prime}}{c_{st}^2  \, |\vec{q} - \vec{k}|^2} 
\nonumber\\
&+& \frac{{\mathcal R}_{\vec{k}}^{\prime} \, {\mathcal R}_{\vec{q} - \vec{k}} }{c_{st}^2  \, k^2}\biggr] + 
\frac{( 2 {\mathcal H}^2 - {\mathcal H}^{\prime}) ({\mathcal H}^2 - {\mathcal H}^{\prime})}{{\mathcal H}^2 \, c_{st}^4 \, k^2 \, |\vec{q}- \vec{k}|^2 } \, {\mathcal R}_{\vec{k}}^{\prime}\,  {\mathcal R}_{\vec{q} - \vec{k}}^{\prime}\biggr\},
\label{EQ14}
\end{eqnarray}
where the explicit dependence of the curvature perturbations upon the conformal time coordinate 
has been dropped for the sake of conciseness (e.g. ${\mathcal R}_{\vec{k}}(\tau) \to {\mathcal R}_{\vec{k}}$ and so on).
Inserting Eq. (\ref{EQ12}) into Eq. (\ref{EQ14}) and keeping the leading contribution inside 
the sound horizon the effective anisotropic stress becomes:
\begin{eqnarray}
&& \Pi_{ij}(\vec{q},\tau) = - \frac{1}{(2\pi)^{3/2} \, \ell_{P}^2 \, a^2} \int d^{3} k\,k_{i} \, k_{j} \,\, \overline{{\mathcal R}}(\vec{k}\,) \, 
\overline{{\mathcal R}}(\vec{q} -\vec{k}\,) \, M[k\,r_{s}(\tau);\, |\vec{q} - \vec{k}| \, r_{s}(\tau)],
\label{EQ16}\\
&& M[k \,r_{s}(\tau); \, |\vec{q} - \vec{k}| \, r_{s}(\tau)] = \frac{ 2 ({\mathcal H}^2 - {\mathcal H}^{\prime})}{{\mathcal H}^2} 
\frac{\sin{k r_{s}(\tau)}}{[k\, r_{s}(\tau)]^{\alpha}}\,\frac{\sin{[|\vec{q} - \vec{k}| r_{s}(\tau)]}}{[|\vec{q} - \vec{k}| \, r_{s}(\tau)]^{\alpha}}.
\label{EQ17}
\end{eqnarray}
From Eqs. (\ref{EQ16}) and (\ref{EQ17}) the full evolution of each tensor polarization obeys 
\begin{equation}
h_{\lambda}^{\prime\prime} + 2 {\mathcal H} h_{\lambda}^{\prime} + q^2 h_{\lambda} = - 2 \,\ell_{P}^2 \,a^2(\tau) \, \Pi_{\lambda},
\label{EQ18}
\end{equation}
where $\Pi_{\lambda}(\vec{q},\tau) = e_{i\, j}^{(\lambda)}(\hat{q}) \Pi^{i\,j}(\vec{q},\tau)/2$ (with $\lambda  = \oplus, \, \otimes$) is obtained by projecting the result of Eq. (\ref{EQ16}) on each of the two tensor polarizations:
\begin{equation}
\Pi_{\lambda}(\vec{q},\tau) = 
- \frac{1}{2\,(2\pi)^{3/2} \, \ell_{P}^2 \, a^2} \int d^{3} k\,\,k^2  \, s_{\lambda}(\hat{q},\hat{k}) \,\, \overline{{\mathcal R}}(\vec{k}\,) \,\overline{{\mathcal R}}(\vec{q} -\vec{k}\,) \, M[k\, r_{s}(\tau) ;\, |\vec{q} - \vec{k}| \, r_{s}(\tau)],
\label{EQ19}
\end{equation}
and $s_{\lambda}(\hat{q},\hat{k}) = e_{i\, j}^{(\lambda)}(\hat{q}) \,\hat{k}^{i}\, \hat{k}^{j}$. Since the energy density 
of Eq. (\ref{EQ9}) contains both the amplitude $h_{ij}$ and its derivative, the solution of 
Eq. (\ref{EQ18}) can therefore be expressed as:
\begin{eqnarray}
h_{\lambda}(\vec{q},\tau) &=& h^{(1)}_{\lambda}(\vec{q},\tau) - 2 \ell_{P}^2 \int_{\tau_{i}}^{\tau} d \xi \, a^2(\xi) \,G[ q ( \xi- \tau)] \, 
\Pi_{\lambda}(\vec{q}, \xi), 
\label{EQ20}\\
\partial_{\tau} h_{\lambda}(\vec{q},\tau) &=& \partial_{\tau} h^{(1)}_{\lambda}(\vec{q},\tau) - 2 \ell_{P}^2  \int_{\tau_{i}}^{\tau} d \xi \, a^2(\xi) \,\widetilde{\,G\,}[ q ( \xi- \tau)] \, 
\Pi_{\lambda}(\vec{q}, \xi).
\label{EQ21}
\end{eqnarray}
The first-order contributions of Eqs. (\ref{EQ20})--(\ref{EQ21}) coincide with the ones already introduced in Eq. (\ref{EQ6}) and
determine the explicit form of $G[ q ( \xi- \tau)]$ and $\widetilde{\,G\,}[ q ( \xi- \tau)]$ which are are Green's functions of the problem. The correlation function $\langle \Pi_{\lambda}(\vec{q}, \tau)\,  \Pi_{\lambda^{\prime}}(\vec{q\,}^{\prime}, \tau^{\prime}) \rangle$ follows straightforwardly from Eqs. (\ref{EQ13}) and (\ref{EQ19}) and the obtained result determine the power spectrum of the anisotropic stress:
\begin{eqnarray}
\langle \Pi_{i\,j}(\vec{q},\tau) \, \Pi_{m\,n}(\vec{q\,}^{\,\prime},\tau^{\prime}) \rangle &=& \frac{2\pi^2}{q^3} \, P_{\Pi}(q, \tau,\tau^{\prime}) 
{\mathcal S}_{ijmn}(\hat{q}) \delta^{(3)}(\vec{q} + \vec{q}^{\prime}), 
\label{EQ22}\\
\Pi_{i\,j}(\vec{q},\tau) &=& \sum_{\lambda  = \oplus,\,\otimes} e^{(\lambda)}_{i\, j}(\hat{q})  \Pi_{\lambda}(\vec{q},\tau).
\label{EQ22a}
\end{eqnarray}
The power spectrum $P_{\Pi}(q, \tau,\tau^{\prime})$ appearing in Eq. (\ref{EQ22a}) is:
\begin{eqnarray}
P_{\Pi}(q, \tau,\tau^{\prime}) &=& \frac{q^3}{4 \pi \, a^2(\tau) \, a^2(\tau^{\prime}) \ell_{P}^4} \int d^{3} k \, k^4\, (1 - \mu^2)^2
\frac{\overline{P}_{{\mathcal R}}(k)}{k^3} \, \frac{\overline{P}_{{\mathcal R}}(|\vec{q} -\vec{k}|)}{|\vec{q} - \vec{k}|^3}
\nonumber\\
&\times& M[k \,r_{s}(\tau); \, |\vec{q} - \vec{k}| \, r_{s}(\tau)] \,M[k \,r_{s}(\tau^{\prime}); \, |\vec{q} - \vec{k}|  r_{s}(\tau^{\prime})],
\label{EQ23}
\end{eqnarray}
where we defined $\mu = \cos{\vartheta}$ and $d^{3} k = k^2 \, d\mu \, d\varphi \, d k$.
The result of Eq. (\ref{EQ23}) and its descendants have been already obtained in an analog context 
in Ref. \cite{TEN} where, however, the role of the power spectrum of curvature perturbations was played by 
the spectrum of a waterfall field \cite{TENa}. 

Since the first-order tensor amplitudes are not correlated with their curvature counterpart, the expectation values containing a first-order tensor amplitude and a curvature perturbation vanish (i.e.
$\langle h_{ij }(\vec{q},\tau) \, {\mathcal R}_{\vec{p}}(\tau) \rangle =0$). Consequently, when 
Eqs. (\ref{EQ20})--(\ref{EQ21}) are inserted into Eq. (\ref{EQ9}), the average of $\rho_{gw}$  
leads directly to the total spectral energy density in critical units that is given as the sum 
of the first-order term $\Omega_{gw}^{(1)}(q,\tau)$ (already discussed in Eq. (\ref{EQ10}))  and of the second-order contribution $\Omega_{gw}^{(2)}(q,\tau)$
following from Eqs. (\ref{EQ22})--(\ref{EQ23}):
\begin{equation}
\Omega_{gw}^{(2)}(q,\tau) = \frac{q^3}{24 \, \pi\, H^2 \, a^4} \int d^{3} k \, (1- \mu^2)^2 \, \frac{k\, \overline{P}_{\mathcal R}(k)\, \overline{P}_{{\mathcal R}}(|\vec{q} - \vec{k}|) }{|\vec{q} - \vec{k}|^3} \,
  \biggl[ {\mathcal I}^2(\vec{k}, \vec{q}, \tau) + {\mathcal J}^2(\vec{k}, \vec{q}, \tau) \biggr],
\label{EQ24}
\end{equation}
 where the two integrals 
${\mathcal I}(\vec{k}, \vec{q}, \tau)$ and ${\mathcal J}(\vec{k}, \vec{q}, \tau)$ are given by:
\begin{eqnarray}
{\mathcal I}(\vec{k},\vec{q},\tau) &=& \int_{\tau_{i}}^{\tau} a(\xi) \sin[q(\xi - \tau)] \, M[k \,c_{st}\, \xi ;\, |\vec{q} - \vec{k}| \,c_{st} \,\xi]\, d\xi
\nonumber\\
&=& \frac{2( \alpha + 1)}{\alpha\, [ k \tau_{1} c_{st}]^{\alpha} [ |\vec{q} - \vec{k}| c_{st}]^{\alpha}} \int_{\tau_{i}}^{\tau} \frac{\sin{[q(\xi - \tau)]}\, \sin{(k c_{st} \xi)}\, \sin{[|\vec{q} - \vec{k}| c_{st} \xi]}}{\xi^{\alpha}},
\label{EQ25}\\ 
{\mathcal J}(\vec{k},\vec{q},\tau) &=& \int_{\tau_{i}}^{\tau} a(\xi) \cos[q(\xi - \tau)] \, M[k \,c_{st}\, \xi ;\, |\vec{q} - \vec{k}| \,c_{st} \,\xi]\, d\xi
\nonumber\\
&=& \frac{2( \alpha + 1)}{\alpha\, [ k \tau_{1} c_{st}]^{\alpha} [ |\vec{q} - \vec{k}| c_{st}]^{\alpha}} \int_{\tau_{i}}^{\tau} \frac{\cos{[q(\xi - \tau)]}\, \sin{(k c_{st} \xi)}\, \sin{[|\vec{q} - \vec{k}| c_{st} \xi]}}{\xi^{\alpha}}.
\label{EQ26}
\end{eqnarray}
The integrals of Eqs. (\ref{EQ25}) and (\ref{EQ26}) have been presented in the case 
of a constant sound speed and in the regime $q c_{st} \tau \gg 1$, $k c_{st} \tau \gg 1$ and $|\vec{q} - \vec{k}| c_{st} \tau \gg 1$.  For a closed analytical expression the integration over $k$ can be divided in two complementary domains: the region $k > q$ [with cut-off $q_{max}  = {\mathcal O}(\mathrm{MHz})$] and the region $k< q$ [with infrared cutoff $q_{p}= {\mathcal O}(\mathrm{aHz})$]. This strategy has been already employed  in Ref.  \cite{TEN} and it leads to a very good quantitive agreement with the numerical result. 

With these specifications the total spectral energy density in critical units can be expressed in the following 
manner
\begin{eqnarray}
&& h_{0}^2 \Omega_{gw}(q,\tau_{0}) = \frac{h_{0}^2\Omega_{R0} r_{T} {\mathcal A}_{\mathcal R}}{12}\, \biggl( \frac{q}{q_{r}}\biggr)^{2 ( 1 -\alpha)} \, \biggl( \frac{q}{q_{p}} \biggr)^{n_{T}}\biggl[ 1 
+ \frac{32  (1 + \alpha)^2\, {\mathcal A}_{{\mathcal R}}}{15\, \alpha^2\, r_{T}\, | q c_{st} \tau_{0}|^{2\alpha}\, c_{st}^{2\alpha}}  f(m,q)\biggl( \frac{q}{q_{p}}\biggr)^{\gamma}\biggr]
\nonumber\\
&&\,\,\,\,\,f(m, q) = \biggl[\frac{m-2}{(2 m -1) (m+1)} - \frac{1}{m+1} \biggl(\frac{q_{p}}{q} \biggr)^{m+1} + \frac{1}{2m -1} \biggl(\frac{q_{max}}{q}\biggr)^{2 m -1}\biggr], 
\label{EQ27}
\end{eqnarray}
where $m= n_{s} - 2 \alpha$ and $ \gamma = 2 (n_{s} -1) - n_{T} $. Thanks to Eq. (\ref{EQ1}) the multiplicative factors appearing in front of the square bracket in Eq. (\ref{EQ27}) reconstruct  the first-order result 
and  coincide with the expression of Eq. (\ref{EQ10}); the second term inside the squared bracket gives instead the second-order correction. The spectral slopes and the numerical 
factors differ from the ones arising in the radiation-dominated case but 
 Eqs. (\ref{EQ27}) and of Eq. (\ref{EQ10}) have been purposely written in a similar manner. 
 Since $q c_{st} \tau_{0} \gg 1$  the relative correction appearing in Eq. (\ref{EQ27}) 
is always much smaller than $1$ and even smaller than the analog contribution arising 
in Eq. (\ref{EQ10}) in the case of the radiation-dominated evolution. 

For a quantitative estimate of the correction appearing in Eq. (\ref{EQ27}) we note 
that $q_{p}$ and $\tau_{0}^{-1}$ are both in the aHz range so that we can trade 
$\tau_{0}^{-1}$ for $q_{p}$. Equation (\ref{EQ27}) becomes therefore:
\begin{equation}
h_{0}^2 \Omega_{gw}(q,\tau_{0}) = r_{T} \frac{h_{0}^2\Omega_{R0}\,{\mathcal A}_{\mathcal R}}{12}\, \biggl( \frac{q}{q_{r}}\biggr)^{2 ( 1 -\alpha)} \, \biggl(\frac{q}{q_{p}} \biggr)^{n_{T}}\biggl[ 1 
+ \frac{32  \, {\mathcal A}_{{\mathcal R}}}{15\, r_{T}\, c_{st}^{4\alpha}} \biggl(1 + \frac{1}{\alpha}\biggr)^2 f(m,q) \biggl(\frac{q}{q_{p}}\biggr)^{ \gamma -2 \alpha}\biggr].
\label{EQ29}
\end{equation}
Let us now consider some simple numerical estimate of the first- and second-order contributions. As far as the pre-factor is concerned we have that it is of the order of $10^{-16}$. In fact, in the concordance paradigm  $h_{0}^2 \Omega_{R0} = 4.15 \times 10^{-5}$, ${\mathcal A}_{{\mathcal R}} = {\mathcal O}(2.41) \times 10^{-9}$ and $r_{T} \leq {\mathcal O}(0.01)$. The frequency $q_{r}$, depending on the 
model can be of the order of few Hz or even smaller.  For $q$ between $10^{-5}$ Hz and $q_{max} = 100\,$MHz 
$f(m,q) = {\mathcal O}(10^{-2})$ for $0.9 <n_{s} <1$ and $1/2 \leq \alpha < 1$. With the approximate identification $\tau_{0} \simeq q_{p}^{-1}$ the second-order correction in the case of a radiation-dominated universe is recovered, up to numerical factors, for $\alpha =1$. We can therefore estimate that for frequencies between the $\mu$Hz and the MHz that 
the second-order correction induced by a stiff phase is more suppressed by a factor $(q/q_{p})^{-2\alpha}$.
In the case of $c_{st} = 1$ (i.e. $\alpha =1/2$) this factor varies between $10^{-12}$ and $10^{-24}$. A further 
(frequency-independent) suppression is given by ${\mathcal A}_{{\mathcal R}}/r_{T} = {\mathcal O}(10^{-7})$.
Barring therefore for irrelevant numerical factor the second term inside the square bracket in Eq. (\ref{EQ29}) 
ranges between $10^{-21}$ and $10^{-33}$ assuming $0.9 <n_{s} <1$ and $1/2 \leq \alpha < 1$.

While the conclusions based on Eq. (\ref{EQ29}) hold for a generic stiff phase driven by fluid sources,
the more specific case of a scalar field $\varphi$ leads to the same results but for the sake of accuracy the main differences will now be outlined. The evolution 
of the scalar normal modes can be derived from the following effective action 
\begin{equation}
S_{{\mathcal R}} = \frac{1}{2} \int d^{4} x \biggl(\frac{\varphi^{\prime}}{{\mathcal H}}\biggr)^2 \, \sqrt{- \overline{g}} \, \overline{g}^{\alpha\beta} \partial_{\alpha} {\mathcal R} \, \partial_{\beta} {\mathcal R},
\label{EQ30}
\end{equation} 
applying, for instance, in the context of quintessential inflation where 
the stiff phase is driven by the kinetic energy and $\varphi$ coincides with the inflaton-quintessence field \cite{SEVEN,EIGHT}. As before the effective anisotropic stress can be computed from Eq. (\ref{EQ13a}) using the standard Landau-Lifshitz strategy \cite{FOURTEENa}; therefore,
in the longitudinal gauge, the result is
\begin{eqnarray}
\Pi_{\lambda}(\vec{q}, \tau) &=& - \frac{2 ({\mathcal H}^2 - {\mathcal H}^{\prime})}{(2\pi)^{3/2}\,\ell_{P}^2\, a^2(\tau)\,{\mathcal H}^2 } \int\, d^{3}k\, \,k^2\,\, s_{\lambda}(\hat{k}, \hat{q})\, \biggl\{ {\mathcal R}_{\vec{k}} \, {\mathcal R}_{\vec{q} - \vec{k}} + \frac{{\mathcal H}^2 - {\mathcal H}^{\prime}}{{\mathcal H}} \biggl[ \frac{{\mathcal R}_{\vec{k}} \, {\mathcal R}_{\vec{q} - \vec{k}}^{\prime}}{ \, |\vec{q} - \vec{k}|^2} 
\nonumber\\
&+& \frac{{\mathcal R}_{\vec{k}}^{\prime} \, {\mathcal R}_{\vec{q} - \vec{k}} }{\, k^2}\biggr] + 
\frac{( 2 {\mathcal H}^2 - {\mathcal H}^{\prime}) ({\mathcal H}^2 - {\mathcal H}^{\prime})}{{\mathcal H}^2 \,  \, k^2 \, |\vec{q}- \vec{k}|^2 } \, {\mathcal R}_{\vec{k}}^{\prime}\,  {\mathcal R}_{\vec{q} - \vec{k}}^{\prime}\biggr\},
\label{EQ31}
\end{eqnarray}
where $s_{\lambda}(\hat{q},\hat{k})$ has been already introduced above (see Eq. (\ref{EQ19}) and definition thereafter). Equation (\ref{EQ31}) matches  Eq. (\ref{EQ14}) in the limit $c_{st} \to 1$. 
It is interesting to remark that Eq. (\ref{EQ31}) coincides with the result 
 obtained by functional derivation of Eq. (\ref{EQ30}) with respect to the background metric. 
This observation proves, incidentally,  that  the suggestion of Ford and Parker
 \cite{FIFTEEN} (originally formulated for the tensor modes) also applies in the case of the scalar modes. Indeed the effective energy-momentum pseudo-tensor of the 
scalars follows from Eq. (\ref{EQ30}) by considering the curvature perturbations and the background metric as independent variables. By taking the functional derivative with respect to $\overline{g}_{\mu\nu}$ we have that the energy-momentum pseudo-tensor of the curvature inhomogeneities is:
\begin{equation}
\delta S_{{\mathcal R}} = \frac{1}{2} \int d^{4} x \, \sqrt{-\overline{g}}\, {\mathcal T}_{\mu\nu} \, \delta \overline{g}^{\mu\nu}\Rightarrow 
{\mathcal T}_{\mu\nu} = \biggl(\frac{\varphi^{\prime}}{{\mathcal H}}\biggr)^2 \biggl[ \partial_{\mu} {\mathcal R} \partial_{\nu} {\mathcal R} -
\frac{1}{2}\overline{g}_{\mu\nu} \biggl(\overline{g}^{\alpha\beta} \partial_{\alpha} {\mathcal R} \partial_{\beta} {\mathcal R} \biggr)\biggr].
\label{EQ33}
\end{equation}
Recalling that the unperturbed metric must be used to raise and lower the indices, the various components of the effective energy-momentum tensor ${\mathcal T}_{\mu}^{\nu}$ 
are 
\begin{eqnarray}
{\mathcal T}_{0}^{0} = \rho_{{\mathcal R}},\qquad 
{\mathcal T}_{i}^{\,\,\,j} = - p_{{\mathcal R}}\,\,\delta_{i}^{\,\,\,j}  + \Pi_{i}^{\,\,\,\,j},
\label{EQ34}
\end{eqnarray}
where $\rho_{{\mathcal R}}$, $p_{{\mathcal R}}$ and $ \Pi_{i}^{\,\,\,\,j}$ are given, respectively, by:
\begin{eqnarray}
\rho_{{\mathcal R}} &=& \frac{\varphi^{\,\prime \,\,2}}{2 \,{\mathcal H^2} \, a^2}
\biggl[\partial_{\tau} {\mathcal R} \partial_{\tau} {\mathcal R} + \partial_{k} {\mathcal R} \partial_{k} {\mathcal R} \biggr], \qquad p_{{\mathcal R}} = \frac{\varphi^{\,\prime \,\,2}}{2 \,{\mathcal H^2} \, a^2}
\biggl[\partial_{\tau} {\mathcal R} \partial_{\tau} {\mathcal R} - \frac{1}{3} \partial_{k} {\mathcal R} \partial_{k} {\mathcal R} \biggr],
\label{EQ36}\\
 \Pi_{i}^{\,\,\,\,j} &=& - \frac{\varphi^{\,\prime \,\,2}}{{\mathcal H^2} \, a^2} \biggl[\partial_{i} {\mathcal R}\, \partial^{j} {\mathcal R} - \frac{\delta_{i}^{\,\,\,j}}{3} \partial_{k} {\mathcal R} \partial_{k} {\mathcal R} \biggr].
\label{EQ37}
\end{eqnarray} 
Inside the sound horizon (which coincides with the Hubble radius in the case of a single scalar field) 
the spatial gradients are of the same order of the time-derivatives; this means that, from Eq. (\ref{EQ36}), the approximate equation of state is $p_{{\mathcal R}} \simeq \rho_{{\mathcal R}}/3$. 
Equation (\ref{EQ37}) can be translated in Fourier space and projected on the tensor polarizations  $e_{\lambda}^{i\,\, j}(\hat{q})$ (with $\lambda =\oplus,\,\otimes$) so that $\Pi_{\lambda}(\vec{q},\,\tau)$
will be, in this case,
\begin{equation} 
\Pi_{\lambda}(\vec{q},\,\tau) = \frac{1}{2} e^{i}_{j\,\,\lambda}(\hat{q})\, \Pi_{i j}(\vec{q},\tau) =
- \frac{2 ({\mathcal H}^2 - {\mathcal H}^{\prime})}{(2\pi)^{3/2}\,\ell_{P}^2\, a^2(\tau)\,{\mathcal H}^2 } \int\, d^{3}k\, \,k^2\,\, s_{\lambda}(\hat{k}, \hat{q})\, {\mathcal R}_{\vec{k}} \, {\mathcal R}_{\vec{q} - \vec{k}};
\label{EQ38}
\end{equation}
in Eq. (\ref{EQ38}) the scalar field has been eliminated by using the background identity 
$2({\mathcal H}^2 - {\mathcal H}^{\prime}) = \ell_{P}^2 \, \varphi^{\prime\, 2}$.
As anticipated Eqs. (\ref{EQ38}),  (\ref{EQ31}) and (\ref{EQ14}) coincide 
inside the sound horizon: in all three cases the leading-order result is ${\mathcal R}_{\vec{k}} \, {\mathcal R}_{\vec{q} - \vec{k}}$
while all the other terms are quickly negligible when $k \gg a\, H$ and 
$|\vec{q} - \vec{k}| \gg a\, H$.

All in all  the first-order contribution to the total spectral energy density  of the relic gravitons expected from the concordance scenario is, at most,  $h_{0}^2 \Omega^{(1)}_{gw}(q,\tau_{0}) = {\mathcal O}(10^{-16.5})$ for typical frequencies between few $\mu$Hz and the MHz. Here we considered tensor modes exiting the Hubble radius during inflation and reentering when the plasma is already dominated by a stiff plasma. In this case the spectral energy density exhibits a sharply growing branch that depends on the total sound speed of the plasma after inflation. Since the curvature inhomogeneities will also reenter all along the stiff epoch, a secondary contribution to the spectral energy density of the relic gravitons will be induced exactly as in the case of a radiation-dominated plasma but with rather different quantitive features. Given that the curvature perturbations reentering the sound horizon in the stiff phase are generally less suppressed than those crossing the sound horizon when the plasma is dominated by radiation, an enhanced second-order contribution might be expected for the tensor modes induced by the curvature inhomogeneities. Overall this expectation is correct since the sum of the first-order and of the second-order tensor contributions is indeed larger than the analog result valid for curvature inhomogeneities reentering during radiation. However, for a stiff plasma, the first-order contribution  always exceeds the corresponding second-order results that are therefore negligible for all phenomenologically relevant purposes. A byproduct of the present analysis suggests that the effective anisotropic stress of the scalar modes follows by taking the functional derivative of the second-order action of curvature inhomogeneities with respect to the background metric. We remark that inside the sound horizon this strategy gives exactly the same result of other gauge-dependent derivations based on the more standard Landau-Lifshitz approach.

The author wishes to thank T. Basaglia, A. Gentil-Beccot, S. Rohr and J. Vigen of the CERN Scientific Information Service for their kind help.


\newpage

\begin{thebibliography}{99}

\itemsep -3pt

\bibitem{ONE} Ya. Zeldovich, Sov. Phys. Usp. {\bf 6}, 475 (1964) [Usp. Fiz. Nauk. {\bf 80}, 357 (1963)].

\bibitem{TWO} A. D. Sakharov, Sov. Phys. JETP {\bf 22}, 241 (1966) [Zh. Eksp. Teor. Fiz. {\bf 49}, 345 (1965)].

\bibitem{THREE} L.P. Grishchuk, Annals N. Y. Acad. Sci. {\bf 302}, 439 (1977).

\bibitem{FOUR} L. H. Ford, Phys. Rev. D {\bf 35}, 2955  (1987).

\bibitem{FIVE}  B. Spokoiny, Phys. Lett. B {\bf 315}, 40 (1993).

\bibitem{SIX} M.~Giovannini,  Phys.\ Rev.\ D {\bf 58}, 083504 (1998).  

\bibitem{SEVEN}   P.~J.~E.~Peebles and A.~Vilenkin,  Phys.\ Rev.\ D {\bf 59}, 063505 (1999).

\bibitem{EIGHT} M.~Giovannini, Phys.\ Rev.\ D {\bf 60}, 123511 (1999). 

\bibitem{EIGHTa} I.~Antoniadis, A.~Karam, A.~Lykkas and K.~Tamvakis,  JCAP {\bf 1811}, no. 11, 028 (2018).

\bibitem{EIGHTb} V.~M.~Enckell, K.~Enqvist, S.~Rasanen and L.~P.~Wahlman,  JCAP {\bf 1902}, 022 (2019).

\bibitem{EIGHTc}  I.~Antoniadis, A.~Karam, A.~Lykkas, T.~Pappas and K.~Tamvakis,   JCAP {\bf 1903},  005 (2019).

\bibitem{EIGHTd} T.~Tenkanen,  Phys.\ Rev.\ D {\bf 99},  063528 (2019).

\bibitem{EIGHTe} M.~Giovannini, Class. Quant. Grav. {\bf 36}, 235017 (2019).

\bibitem{EIGHTf} M.~Giovannini, Prog. Part. Nucl. Phys. {\bf 112}, 103774 (2020).

\bibitem{EIGHTg} J.~de Haro, J.~Amor\'os and S.~Pan, Phys. Rev. {\bf 94}, 064060 (2016); Eur. Phys. J. C {\bf 79}, 
 505 (2019).

\bibitem{EIGHTh} K.~Dimopoulos and L.~Donaldson-Wood, Phys. Lett. B {\bf 796}, 26 (2019); 
E.~I.~Guendelman, R.~Herrera and P.~Labrana, [arXiv:2005.14151 [gr-qc]]; J.~Haro and L.~Arest\'e Sal\'o, [arXiv:2004.11843 [gr-qc]].

\bibitem{EIGHTi} A.R. Liddle, S.M. Leach, Phys. Rev. D {\bf 68}, 103503 (2008); M. Giovannini, Phys. Lett. B {\bf 668}, 44 (2008);
Class. Quant. Grav. \textbf{26}, 045004 (2009).

\bibitem{NINE} K.~N.~Ananda, C.~Clarkson and D.~Wands,  Phys.\ Rev.\ D {\bf 75}, 123518 (2007).

\bibitem{TEN}  M.~Giovannini, Phys.\ Rev.\ D {\bf 82},  083523 (2010).

\bibitem{TENa}  D.~H.~Lyth,  Prog.\ Theor.\ Phys.\ Suppl.\  {\bf 190}, 107 (2011);
 S.~Clesse,  Phys.\ Rev.\ D {\bf 83}, 063518 (2011).
 
\bibitem{THIRTEENa} S. Y. Khlebnikov and I. I. Tkachev, Phys. Rev. D {\bf 56}, 653 (1997).

 \bibitem{FOURTEEN} S.~Antusch, F.~Cefala and S.~Orani,  JCAP {\bf 1803}, 032 (2018).
 
\bibitem{FOURTEENa} L. D. Landau and E. M. Lifshitz, {\it The Classical Theory of Fields}, (Pergamon Press, New York, 1971). 
 
\bibitem{FIFTEEN} L. H. Ford and L. Parker, Phys. Rev. {\bf D16}, 1601 (1977).

 \bibitem{NINEa}  V.~N.~Lukash,  Sov.\ Phys.\ JETP {\bf 52}, 807 (1980) [Zh. Eksp. Teor. Fiz. {\bf 79}, 1601 (1980)].

\bibitem{ELEVEN} J.~Hwang, D.~Jeong and H.~Noh, Astrophys.\ J.\  \textbf{842}, 46 (2017).

\bibitem{TWELVE}  R.~g.~Cai, S.~Pi and M.~Sasaki,  Phys.\ Rev.\ Lett.\  {\bf 122}, 201101 (2019).  

\bibitem{THIRTEEN} M.~Giovannini, arXiv:2005.04962 [hep-th].

\end{thebibliography}
\end{document}